%% file: CCDMC.tex
\documentclass[aps,prl,showpacs,twocolumn,letterpaper,superscriptaddress,nofootinbib]{revtex4-1}
\usepackage{amsmath, amssymb,bm}
\usepackage{natbib}
\usepackage{graphicx}
\usepackage{epsfig}
\usepackage{dcolumn}
\newcolumntype{d}[1]{D{.}{.}{#1}}
\usepackage{array}
\usepackage{tabularx}
\newcolumntype{L}[1]{>{\hsize=#1\hsize\raggedright\arraybackslash}X}
\newcolumntype{R}[1]{>{\hsize=#1\hsize\raggedleft\arraybackslash}X}
\newcolumntype{C}[1]{>{\hsize=#1\hsize\centering\arraybackslash}X}

\newcommand{\la}{\langle} 
\newcommand{\ra}{\rangle}
\newcommand{\be}{\begin{equation}} 
\newcommand{\ee}{\end{equation}}
\newcommand{\ba}{\begin{eqnarray}} 
\newcommand{\ea}{\end{eqnarray}}
\newcommand{\ad}{a^\dagger} 

\newcommand{\ve}{\varepsilon} 
\newcommand{\gs}{|\Psi_{\rm gs} \rangle}

\newcommand{\Dt}{\Delta \tau} 
\newcommand{\bn}{{\bf n}}
\newcommand{\bbm}{{\bf m}}
\newcommand{\mh}{\mathcal{H}}

\newcommand{\mS}{\mathcal{S}} 
\newcommand{\mE}{\mathcal{E}}
\newcommand{\mP}{\mathcal{P}}
\newcommand{\mN}{\mathcal{N}}

\newcommand{\ssec}[1]{\emph{#1}.---}

\begin{document}


\title{Quantum Monte Carlo with Coupled-Cluster wave functions}

\author{Alessandro Roggero}
\email[]{roggero@science.unitn.it}
\affiliation{Physics Department, University of Trento, via Sommarive 14, I-38123 Trento, Italy}
\affiliation{INFN-TIFPA, Trento Institute for Fundamental Physics and Applications}

\author{Abhishek Mukherjee}
\email[]{mukherjee@ectstar.eu}
\affiliation{ECT*, Villa Tambosi, I-38123 Villazzano (Trento), Italy}
\affiliation{INFN-TIFPA, Trento Institute for Fundamental Physics and Applications}

\author{Francesco Pederiva}
\email[]{pederiva@science.unitn.it}
\affiliation{Physics Department, University of Trento, via Sommarive 14, I-38123 Trento, Italy}
\affiliation{INFN-TIFPA, Trento Institute for Fundamental Physics and Applications}


\date{\today}

\begin{abstract}
We introduce a novel many body method which combines two powerful many body techniques, viz., 
quantum Monte Carlo and coupled cluster theory. Coupled cluster wave functions are introduced
as importance functions 
in a Monte Carlo
method designed for the configuration interaction framework to provide rigorous upper bounds to 
the ground state energy. We benchmark our method on the homogeneous electron gas in momentum space.
The importance function used is the coupled cluster doubles wave function. 
We show that the computational resources required
in our method scale polynomially with system size. Our energy upper bounds are in very
good agreement with previous calculations of similar accuracy,
and they can be systematically improved by including higher order excitations in the coupled cluster wave function.    
\end{abstract}


\maketitle
\input{introduction}

\input{method}

\input{results}

\input{conclusions}

\begin{acknowledgments}
Computations were performed at the Open Facilities at Lawrence Livermore National Laboratory.
We would like to thank C.~Umrigar for stimulating discussions and careful reading of the manuscript, 
and Y.~Alhassid for comments on the manuscript.
\end{acknowledgments}

\bibliography{CCDMC}

\end{document}

%% file: introduction.tex
\ssec{Introduction}Quantum Monte Carlo (QMC) methods have now become standard tools
for computations in a wide variety of strongly correlated systems \cite{Nightingale1998book} 
ranging from quantum chemistry \cite{Ceperley1995, *Anderson2007} to condensed matter \cite{Binder1995, *Foulkes2001}, 
and nuclear physics \cite{Pieper2005}. The most celebrated capability of such methods is to provide very accurate
estimates of ground state and thermodynamic properties of many-body systems, while enjoying a very favorable scaling
of computational time with system size.

In fermionic systems, QMC methods are plagued by the `sign problem'. 
In principle, the sign problem can be circumvented in the `fixed-node' (FN) approximation with the help
of importance sampling with a trial ground-state wave function \cite{Reynolds1982} . 
 The spectacular success of the FN QMC is largely because of the development of high quality
trial ground-state wave functions. However, most of these wave functions have forms which are convenient for calculations
in coordinate space ($r$-space).
 
The lack of accurate and computationally efficient trial ground-state wave functions has, thus far, 
precluded a wide exploration of these algorithms
 within the configuration interaction (CI) scheme (see, however, Ref.~\onlinecite{Kolodrubetz2012}). 
Generally, QMC methods within the CI scheme tend to rely on auxiliary fields introduced 
via the Hubbard-Stratonovich transformation \cite{Sugiyama1986, *Zhang2004, *PhysRevA.78.023625,
  *PhysRevA.85.051601, *PhysRevA.84.061602}.
Much recent interest, however, was sparked by the demonstration that even within the CI scheme it is possible to
apply stochastic projection to systems much larger than what would be possible using conventional matrix diagonalization
\cite{Booth2013}.
 
The development of an efficient ground state QMC algorithm within the CI scheme/momentum space would be of great interest 
in nuclear physics, where most of the modern interactions are written in non-local forms \cite{VanKolck1999}; 
and in electronic structure calculations with non-local pseudopotentials \cite{Pickett1989}.

On the other hand, coupled cluster (CC) wave functions provide very accurate and size extensive approximations for 
the ground state wave function for many physical systems of interest \cite{Kummel2002}. The CC energies  
towards accurate QMC calculations in many-body systems \cite{Pedersen2011}.  In fact, CC calculations 
are considered to be the `gold standard' in quantum chemistry \cite{Bartlett2007}. More recently, CC theory has also 
reemerged as a method of choice in nuclear structure calculations \cite{Kowalski2004, *Hagen2007}.  

In this Letter, we introduce a novel scheme to combine these two powerful many-body techniques. We propose to use CC wave functions
as importance functions for a projection quantum Monte Carlo algorithm within the configuration interaction scheme,
which we call configuration interaction Monte Carlo (CIMC). In CIMC the ground state wave function of a CI Hamiltonian is filtered
out by propagating the amplitudes of an initial arbitrary wavefunction via a random walk in the many body Hilbert space spanned 
by a basis of Slater determinants \cite{Mukherjee2013}. 
The CC wave functions are used to guide this random walk via importance sampling in order to circumvent the 
sign problem. Our method provides rigorous upper bounds to the ground state energy whose tightness 
can be systematically improved by including higher order excitations in the CC wave function.
 
We apply our method to the three dimensional homogeneous electron gas (3DEG) in momentum space. 
The 3DEG is described by a simple Hamiltonian;
it nevertheless encapsulates many of the difficulties associated with modern many-body theories. In particular,
it has both the weakly and strongly correlated regimes which can be accessed via a single tunable density parameter,
 the Wigner-Seitz radius $r_s$, thus providing an ideal system for benchmaking many body theories 
\cite{Ceperley1980, Perdew1992, Kwon1998, Rios2006, Shepherd2012b, Shepherd2012c}.

%% file: method.tex
\ssec{Monte Carlo in Fock space} The CIMC projection algorithm is discussed in detail in Ref~\onlinecite{Mukherjee2013}. 
We describe it briefly for completeness.
Consider a general second quantized fermionic CI Hamiltonian which includes only two-body interactions
(atomic units will be used throughout)
\be
H=\sum_{i \in \mS} \ve_i \ad_i a_i + \sum_{abij \in \mS} V^{ab}_{ij}\ad_a \ad_b a_i a_j \;,
\label{eqham}
\ee
where $\ad_i$ creates a particle in the single-particle (sp)
state labeled by $i$ ($i$ is a collective label for all sp quantum numbers).  
The set $\mS$ of sp states is assumed to be finite and of size ${\cal N}_s$.
The $V^{ab}_{ij}$ are general two-body interaction matrix elements. 

For homogeneous systems we can use the plane wave states, with definite momentum and spin, 
as the sp basis set. The  sp energies are $\ve_i = \mathbf{k}_i^2/2m$, where $\mathbf{k}_i$ 
is the momentum of the $i$-state, and $m$ is the fermion mass.
We include in our sp basis all single-particle states $i$ with $k^2_i \leq k^2_{\rm max}$. In principle, 
cutoff independent results can be obtained by performing  successive
calculations with increasing $k_{\rm max}$ and then extrapolating to $k_{\rm max} \to \infty$. 

For the 3DEG, the $V^{ab}_{ij}$  are given by 
\be
V^{ab}_{ij} = (1 - \delta_{\mathbf{k}_a - \mathbf{k}_i,0}) \delta_{\mathbf{k}_i + \mathbf{k}_j,\mathbf{k}_a+\mathbf{k}_b} \frac{4 \pi}{\Omega} \frac{1}{(\mathbf{k}_a - \mathbf{k}_i)^2}\;. 
\label{eqint}
\ee
The volume $\Omega$  of the simulation box (and hence the unit spacing in momentum) is determined by the density 
and the number of particles $N$ in the simulation. We ignore
the Madelung term because it does not affect the correlation energy. 

The many-body Hilbert space is spanned by the set of all $N$-particle Slater determinants constructed 
from the sp orbitals $i \in \mS$. We will denote these Slater determinants or `configurations' with the 
vector notation,  $|\bn \ra$, where $\bn \equiv \{ n_i\}$ and $n_i = 0,1$ is the occupation number 
of the sp orbital $i$ in $| \bn \ra$. 

The ground state wave function $\Psi_{\rm gs}$ of $H$ can be projected out by repeated applications of the projection 
operator $\mP = 1 - \Dt (H-E_T)$, on some initial wave function $\Psi_0$ which has a non-zero overlap 
with $\Psi_{\rm gs}$,  
\be 
\gs = \lim_{M \to \infty} \mP^{M}|\Psi_0 \ra \;.
\label{eqpro}
\ee
where the energy shift $E_T$  and the imaginary time step $\Dt$ are related by
$\Dt < 2/(E_{\rm max} - E_T)$, with $E_{\rm max}$ being the maximal eigenvalue of $H$ \cite{Trivedi1990}.

In a Monte Carlo algorithm the wave function at any time step $M$, $\Psi_M$, is represented as an ensemble of configurations. 
The one time-step projection (single application of $\mP$) is performed independently and stochastically for each configuration, 
by interpreting the matrix elements  $\la \bbm |\mP| \bn \ra$ as probabilites. 

In general (the 3DEG inclusive), $\mP$ will have one or more negative off-diagonal matrix elements, which cannot be 
interpreted as probabilities. In principle, a stochastic evolution
can still be carried out by evolving `signed' configurations, but this leads to an exponential decay in the signal to noise
ratio with increasing time-step $M$. This is a manifestation of the sign problem. Recently, it was 
shown that the sign problem can be somewhat moderated by including a configuration-annihilation step in the evolution \cite{Booth2009,*Cleland2010}. 
Nevertheless, even the latter algorithm has exponential scaling with the system size, though with a reduced exponent.

In CIMC, we circumvent the sign problem with the help of an importance function $\Phi_G$.   
We define a family of Hamiltonians, $\mh_{\gamma}$, whose off-diagonal matrix
elements are given by \cite{TenHaaf1995,Sorella2000},
\be
  \label{mh1}
  \langle \bbm |\mh_{\gamma} |\bn \rangle  =\left \{ \begin{array}{rr} -\gamma \langle \bbm| H |\bn \rangle  & \mathfrak{s}(\bbm,\bn)  > 0 \\
  \langle \bbm| H |\bn \rangle   &  \mbox{otherwise} \end{array} \right . \;,
\ee
while the diagonal matrix elements are given by,
\be
\label{mh2}
\langle \bn | \mh_{\gamma} | \bn \rangle = \langle \bn | H | \bn \rangle+ (1+\gamma) \displaystyle \sum_{\stackrel{ \bbm \neq
\bn}{\mathfrak{s}(\bbm,\bn) > 0}} \mathfrak{s} (\bbm,\bn) \;.
\ee
where, $\mathfrak{s}(\bbm,\bn) = \Phi_G(\bbm) \langle \bbm | H |\bn \rangle / \Phi_G(\bn)$.

In addition, we define a new propagator $\mP_{\gamma}$ as,
\be
\label{mpg}
\langle \bbm | \mP_{\gamma} |\bn \rangle = 1 - \Dt \Phi_G (\bbm) \langle \bbm | \mh_{\gamma} - E_T|\bn \rangle/  \Phi_G(\bn)\;.
\ee
The propagator $\mP_{\gamma}$, by construction, is free from the sign problem  for $\gamma \geq 0$, and  
filters out the wave function $\Phi_G(\bn) \Psi_{\gamma}(\bn)$, where
$\Psi_{\gamma}(\bn)$ is the ground state wave function of $\mh_{\gamma}$.

The ground state energy $\mathcal{E}_{\gamma}$ of $H_{\gamma}$ is
an strict upper bound for the ground state energy $E_{\rm gs}$ of the true Hamiltonian  $H$ for
$\gamma \geq 0$, and this upper bound is tighter than the variational upper bound
$\langle \Phi_G |H|\Phi_G \rangle$ \cite{TenHaaf1995,Sorella2000,Mukherjee2013}. 
In addition, a linear extrapolation of $\mE_{\gamma}$ from any two values of $\gamma$ to $\gamma = -1$ also provides
a rigorous upper bound on $E_{\rm gs}$ \cite{Beccaria2001}. We found the extrapolation using $\gamma=0,1$ to be best compromise
between accuracy and statistical error. Thus, the  tightest upper bound for the ground state energy $E_0$  is 
$E_{\rm CIMC}=2\mE_{\gamma = 0} - \mE_{\gamma = 1}$.

The simple projection algorithm described above becomes extremely inefficient for large $E_{\rm max}$, i.e, for large $N$ or $\mathcal{N}_s$.
In practice, we use a  much more efficient algorithm (free from any time-step error) which was proposed in Ref.~\onlinecite{Trivedi1990} 
(see also, Ref.~\onlinecite{Sorella2000}). 

\ssec{Coupled cluster importance functions}A good choice for the importance function $\Phi_G$ is crucial for the success
of our method. We need $\Phi_G$ to be sufficiently flexible to include the dominant
correlations in the system, and yet be quick to evaluate. 
In many strongly correlated systems, CC wave functions fulfill the first criterion. 
Below, we describe a recursive algorithm which can be used to evaluate CC wave functions 
very efficiently.

The CC wave function can be written as 
\be
|\Phi_{\rm CC}\ra = e^{\hat{T}} |\Phi_0 \ra
\ee
where $\Phi_0$ is a model wave function and $\hat{T}$ is an operator which generates excitations on $\Phi_0$.
We choose $\Phi_0=\Phi_{\rm HF}$, i.e., the Hartree-Fock wave function. Due to momentum conservation, 
in a homogeneous system the simplest non-trivial approximation for $\hat{T}$ is 
$\hat{T} \equiv \hat{T}_2=\sum_{ij,ab} t^{ab}_{ij}\ad_a \ad_b a_j a_i$, where sp states
$i$ and $j$ ($a$ and $b$) are occupied (unoccupied) in $\Phi_{\rm HF}$. This constitutes the so-called coupled cluster doubles
(CCD) approximation for the ground state wave function. 

Writing the wave function in terms of the excitations on top of $\Phi_{\rm HF}$ it can be shown that the only non-trivial components
of the wave function are given by 
\begin{widetext}
\be
\Phi_{\rm CCD}^{m} \left(  \begin{smallmatrix} p_1 p_2 \dotsm p_{m}  \\ h_1 h_2 \dotsm h_{m} \end{smallmatrix} \right) = \displaystyle \sum^{m}_{\gamma=2} \sum^{m}_{\mu < \nu} (-)^{\gamma + \mu + \nu} t^{p_{\mu} p_{\nu}}_{h_1 h_{\gamma}} \Phi^{m-2}_{\rm CCD} \left( \begin{smallmatrix} p_1 p_2 \dotsm p_{\mu - 1} p_{\mu + 1} \dotsm p_{\nu-1} p_{\nu+1} \dotsm p_{m}  \\ h_2 \dotsm h_{\gamma - 1} h_{\gamma + 1} \dotsm h_{m} \end{smallmatrix} \right) \;,
\label{eqrec}
\ee
\end{widetext}
for even $m > 0$.
In the above equation, $\Phi_{\rm CCD}^{m} \left(  \begin{smallmatrix} p_1 p_2 \dotsm p_{m}  \\ h_1 h_2 \dotsm h_{m} \end{smallmatrix} \right) = \Phi_{\rm CCD} (\bn)$ 
for $| \bn \ra = \ad_{p_1}\ldots \ad_{p_m} a_{h_1} \ldots a_{h_m} |\Phi_{\rm HF}\ra$, with $p_1 < \ldots < p_m$ and $h_1 < \ldots < h_m$. The wave function has a vanishing component for odd $m$, and the component for $m=0$ is fixed by our choice of normalization to be $1$.

Using Eq.~(\ref{eqrec}) we can easily calculate $\Phi_{\rm CCD}$ recursively. The computational
effort for a single calculation has a combinatorial scaling with the average
number of particle-hole excitations in $\Phi_G$. However, it \emph{does not depend}
on $N$ or $\mathcal{N}_s$. 

Although for this exploratory study we have used the CCD wave function, it is evident that similar 
recursive relations can be easily written for other CC type wave functions, e.g., CCSD or CCSDT, 
which will become necessary for applying our method to inhomogeneous systems or, systems with stronger correlations
or many-body interactions.

%% file: results.tex
\ssec{Results for the homogeneous electron gas}In Table~\ref{tabcom} we show the CCD energies, calculated using 
conventional CC theory, along with
the corresponding Monte Carlo energies of an 3DEG system with $N=14$  
and $\mathcal{N}_s = 342$, for $r_s =0.5,1.0$ and $2.0$. We see that, for $r_s=0.5$ the CCD energy is very close to
the corresponding Monte Carlo energy. But, for $r_s =1.0$ and $2.0$ they are, in fact, lower than the corresponding
Monte Carlo energies. Since, $E_{\rm CIMC} \leq \la \Phi_{\rm CCD} |H| \Phi_{\rm CCD}\ra$ (see discussion above), 
this shows, once again, the non-variational nature of the energies obtained from conventional CC theory
\cite{Voorhis2000,*Cooper2010}.

\begin{table}[htbp]
\begin{ruledtabular}
\begin{tabular}{@{\extracolsep{\fill}} d{1}d{8}d{8}|d{8}d{8}}
                             &\multicolumn{4}{c}{Correlation energy (a.u.)} \\
\multicolumn{1}{c}{$r_s$}                & \multicolumn{1}{c}{CCD} & \multicolumn{1}{c}{+ CIMC} & \multicolumn{1}{c}{CCD(1)} & \multicolumn{1}{c}{+ CIMC} \\ 
\hline
\noalign{\smallskip}
        0.5  & -0.572682    & -0.5729(3)   & -0.659641 & -0.5733(2) \\ 
	1.0  & -0.506701    & -0.5021(3)   & -0.657347 & -0.5025(2) \\ 
	2.0  & -0.417946    & -0.40317(2)  & -0.665071 & -0.4029(3)  \\ 
\end{tabular}
\end{ruledtabular}
\caption{ Correlation energies for $N=14$ and $\mathcal{N}_s=342$ from conventional CC theory with 
the CCD and CCD(1) wave functions, along with the corresponding CIMC energies using each as importance 
functions. The numbers in parenthesis indicate statistical error in the last significant digit. }
\label{tabcom}
\end{table}
In principle, the CC amplitudes $t^{ab}_{ij}$ can be obtained from conventional
 CC theory. However, solving the CC equations is computationally expensive. Therefore, we investigated
the possibilty of using less computationally demanding ways of obtaining the $t^{ab}_{ij}$, while still
preserving the structure of the CC wave function. 

A simple option is to compute $t^{ab}_{ij}$ by the second order M\o ller-Plesset perturbation theory (MP2). 
If the CC equations are solved iteratively, then this is equivalent to stopping after the first iteration.
In Table~\ref{tabcom} we also compare the energies obtained from our CIMC projection 
using the MP2, denoted by CCD(1). 

The CCD(1) amplitudes produce a worse approximation to the ground state wave function as compared
to the full CCD amplitudes. Nevertheless, when used as importance functions in our CIMC algorithm, the final estimate for the ground
state energy for both cases are very close. The statistical errors are comparable for $r_s=0.5$ and $1.0$.
For $r_s=2.0$ they are about an order of magnitude lower when the full CCD amplitudes are used.     

The above observation is extremely encouraging because it means that, for our purposes, it may not necessary to 
solve the full CC equations to get reasonable $t^{ab}_{ij}$ amplitudes. Of course, we do not expect the CCD(1) amplitudes 
to be satisfactory for more strongly correlated systems. Still, even in those cases one can, presumably, use computationally
inexpensive approximations to the full CC equations. For the rest of this work, all the CIMC results have been computed using the CCD(1) amplitudes. 

\begin{figure*}[htbp]
\includegraphics[width=\textwidth]{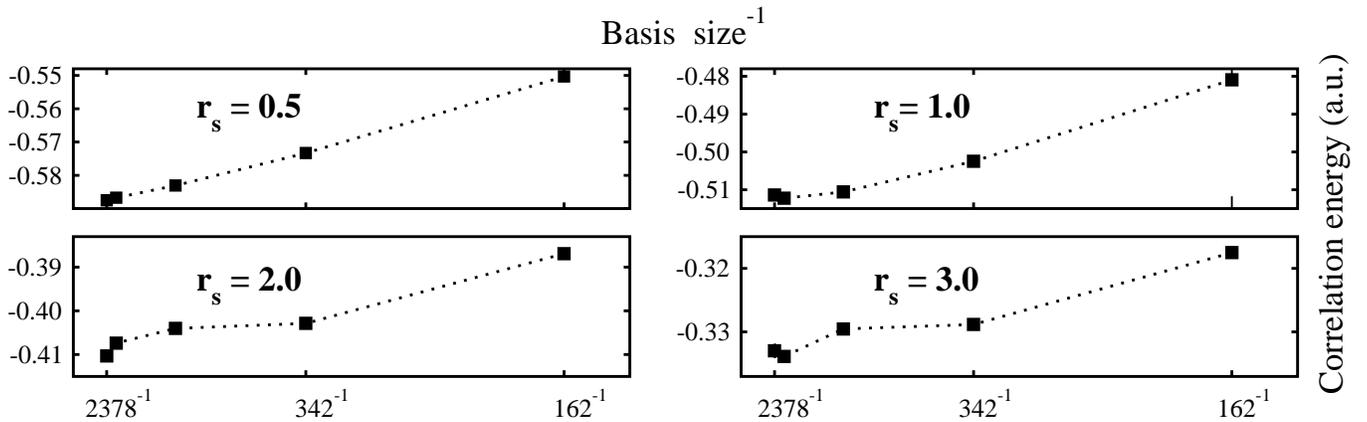}
\caption{Correlation energies for $N=14$ and $r_s=0.5,1.0,2.0$ and $3.0$ as a function of the single particle 
basis size  from CIMC with the CCD(1) importance function. The statistical errors are smaller
than the size of the symbols. The lines are drawn as a guide to the eye.}
\label{figbas}
\end{figure*}
In Fig.~\ref{figbas} we show the CIMC ground state energy estimates for $N=14$ and $r_s = 0.5, 1.0, 2.0$ and $3.0$
for some of our large basis size calculations. In Refs.~\onlinecite{Shepherd2012a, Shepherd2012b, Shepherd2012c} it was suggested 
that for the 3DEG it might be possible to extrapolate to the $\mathcal{N}_s \to \infty$ limit by exploiting a linear $1/{\mathcal{N}_s}$
dependence of the correlation energy for large but finite $\mathcal{N}_s$. Although, for $r_s = 0.5$ and $N=14$  such a 
linear trend in the correlation energy is visible, for the other values of $r_s$ shown in the figure,
no such trend is evident. The situation is similar for calculations we performed with $N=32$ and $54$. Thus, at least up to our largest
basis size $\mathcal{N}_s =2378$, we cannot safely do an extrapolation to $\mathcal{N}_s \to \infty$ with a reasonably low $\chi^2$.

\begin{table}[htbp]
\begin{ruledtabular}
\begin{tabular}{@{\extracolsep{\fill}}d{1} r d{8} d{13}}
    &            & \multicolumn{2}{c}{Correlation energy (a.u)} \\
\noalign{\smallskip}
r_s   & N    &   \multicolumn{1}{c}{CIMC}    & \multicolumn{1}{c}{Other}  \\
\hline 
\noalign{\smallskip}
0.5     & 14   & -0.5875(6) & -0.5959(7)  \mbox{ \cite{Shepherd2012c}} \\
        & 38   & -1.809(4)  & -1.849(1)   \mbox{ \cite{Shepherd2012c}}  \\
        & 54   & -2.354(2)  & -2.435(7)   \mbox{ \cite{Shepherd2012c}} \\
        &      &            & -2.387(2)   \mbox{ \cite{Rios2006}} \\
\noalign{\smallskip}
1.0     & 14   & -0.5114(5) & -0.5316(4)  \mbox{ \cite{Shepherd2012c}}  \\
        & 38   & -1.521(4)  & -1.590(1)   \mbox{ \cite{Shepherd2012c}} \\
        & 54   & -2.053(4)  & -2.124(3)   \mbox{ \cite{Shepherd2012c}} \\
        &      &            & -2.125(2)   \mbox{ \cite{Rios2006}} \\
\noalign{\smallskip}
2.0     & 14   & -0.4103(6) & -0.444(1)   \mbox{ \cite{Shepherd2012c}} \\
        & 38   & -1.134(7)  & -1.225(2)   \mbox{ \cite{Shepherd2012c}} \\
3.0     & 14   & -0.333(1)  &               \\            
\noalign{\smallskip}
\end{tabular}
\end{ruledtabular}
\caption{Correlation energies from CIMC with the CCD(1) importance function for different $r_s$ and $N$
and $\mathcal{N}_s = 2378$. For comparison we have also included the results from Ref.~\onlinecite{Rios2006} 
(basis set independent) and Ref.~\onlinecite{Shepherd2012c} (extrapolated using single point extrapolation from 
$\mathcal{N}_s=1850$ for $N=14$ and from $\mathcal{N}_s = 922$ for $N=38,54$). The numbers in parenthesis
indicate the statistical error in each case.}
\label{tabeng}
\end{table}
In Table~\ref{tabeng} we 
show the ground state energy of different $r_s$ and $N$ for the largest $\mathcal{N}_s (=2378)$ calculated by us. For comparison, we 
also include the energy estimates from Refs.~\onlinecite{Rios2006} and \onlinecite{Shepherd2012c}. 

The energies in Ref.~\onlinecite{Rios2006} are calculated using the $r$-space diffusion Monte Carlo method with an importance function
that included backflow correlation on top of the Slater-Jastrow wave function. These are strict energy upper bounds with a bias
due to the fixed-node approximation. Nevertheless, they are believed to be highly accurate. 
 
On the other hand, in Ref.~\onlinecite{Shepherd2012c} the energies are calculated in a finite CI like basis set (as in this work), using  
the initiator full configuration interaction quantum Monte Carlo method ($i$-FCIQMC). The $\mathcal{N}_s \to \infty$ results are obtained
by using the so called `single point extrapolation' from much smaller values of $\mathcal{N}_s$ than ours. 

Our finite-basis set results are already in good agreement with the other calculations, capturing between $93\%$ to $99\%$ of the 
correlation energy. The energy upper bounds can be systematically improved by including higher order excitations (triples) in the 
CC wave function and by using larger basis sizes.  Possibly, a much faster way to achieve basis set convergence is to use a finite 
basis renormalized Coulomb interaction \cite{Pedersen2011}.

Our method shares many similarities with the FCIQMC method. The FCIQMC method, in principle, can provide exact ground state energies 
for a CI Hamiltonian. However, most calculations are performed using the initiator approximation ($i$-FCIQMC) with adds a bias to the
energy estimate. The energies in $i$-FCIQMC are not necessarily upper bounds to the true ground state energy.
Due to the sign problem, the computational resources required  in either FCIQMC or $i$-FCIQMC scale as the size 
the many body Hilbert space, i.e., they are exponential in the system size, $N$, and the basis size, $\mathcal{N}_s$.
\begin{table}[htbp]
\begin{ruledtabular}
\begin{tabular}{@{\extracolsep{\fill}}d{1}rrrr}
    &     &          & \multicolumn{2}{c}{Computational time (cpu hours)} \\
\multicolumn{1}{c}{$r_s$} & $N$  &$\mathcal{N}_s$ & CIMC &$i$-FCIQMC \\
\hline
\noalign{\smallskip}
0.5 & 14 & 1850    & 384  & 200           \\ 
1.0 & 14 & 1850    & 768  & 2500          \\ 
2.0 & 14 & 1850    & 768  & 2500          \\ 
2.0 & 38 & 922     & 4608 & 16000          \\
\end{tabular}
\end{ruledtabular}
\caption{Computational cost of our method (CIMC) compared with the $i$-FCIQMC method \cite{Shepherd2012c} for different $r_s$, $N$ and $\mathcal{N}_s$.}
\label{tabtim}
\end{table}

Our method, instead, provides strict energy upper bounds, the tightness of which can be systematically improved 
by improving the quality of the importance function. 
Importantly, the computational cost of our 
Monte Carlo algorithm nominally grows as $N^2(\mN_s-N)$, for both computational time (per Monte Carlo step) and memory requirements.
Due to this polynomial scaling, we expect our method to be applicable to much larger systems than those manageable
 by conventional diagonalization methods or by \mbox{($i$-)FCIQMC}. We see evidence of this in Table~\ref{tabtim}, where we compare 
computational time in our method with that in $i$-FCIQMC. The statistical error for the two methods are comparable in each case.

%% file: conclusions.tex
\ssec{Conclusion}
We have introduced a many body technique which combines the power of ground state projection Monte Carlo with 
coupled cluster theory. As a first benchmark and application we studied the homogeneous electron gas in momentum space for large 
single particle basis sizes. Our results are already in very good agreement with existing accurate calculations, and 
they can be systematically improved. In principle, the fixed-node bias can be removed by performing a further projection starting from our CIMC 
wave functions with the true projector $\mP$ and signed configurations using FCIQMC and the semistochastic projector Monte Carlo method \cite{Petruzielo2012}.

The use of CC wave functions as importance functions in our Monte Carlo algorithm serves dual purposes. On the one hand, given that 
the CC wave functions are known to be extremely accurate, they can serve as the prototype for accurate importance functions in Fock 
space Monte Carlo for normal Fermi systems, analogous to the high quality importance functions in $r$-space Monte Carlo \cite{Umrigar2007, *Neuscamman2012}.
 
On the other hand, the energies obtained from conventional CC theory are \emph{not variational}. 
However, the energies we obtain from CIMC are rigorous upper bounds 
for the exact ground state energy. By using CC wave functions in our Monte Carlo we are providing the `best' 
variational energy one can hope to get using the CC class of wave functions. Given the importance of CC type 
wave functions in modern quantum many body calculations, the second purpose is as important as the first.